\documentclass[hidelinks,onefignum,onetabnum]{siamart220329}

\usepackage{lipsum}
\usepackage{amsfonts}
\usepackage{graphicx}
\usepackage{epstopdf}
\usepackage{amsmath}
\usepackage{listings}
\usepackage{algpseudocode}
\usepackage{placeins}
\ifpdf
  \DeclareGraphicsExtensions{.eps,.pdf,.png,.jpg}
\else
  \DeclareGraphicsExtensions{.eps}
\fi

\newsiamremark{remark}{Remark}
\newsiamremark{hypothesis}{Hypothesis}
\crefname{hypothesis}{Hypothesis}{Hypotheses}
\newsiamthm{claim}{Claim}

\headers{Implicitly Restarted Lanczos}{P.S. Negi and C. Arratia}

\title{Structure preserving restarts of the non-symmetric Lanczos algorithm via the implicitly shifted LR algorithm\thanks{Submitted to the editors DATE.
\funding{This work was funded by the Nordita the Swedish Research Council Grant No. 2018-04290. Nordita is partially supported by Nordforsk.}}}

\author{P. S. Negi\thanks{Nordita, Stockholm University, KTH Royal Institute of Technology, Stockholm, Sweden  (\email{prabal.negi@su.se}).}
\and C. Arratia\thanks{Nordita, Stockholm University, KTH Royal Institute of Technology, Stockholm, Sweden (\email{cristobal.arratia@su.se}).}
}

\usepackage{amsopn}

\ifpdf
\hypersetup{
  pdftitle={Implicitly restarted Lanczos},
  pdfauthor={P.S. Negi and C. Arratia}
}
\fi

\newcommand{\refeqn}[1]{equation~(\ref{#1})}

\begin{document}

\maketitle

\begin{abstract}
The implicitly shifted QR iteration is used as a restart procedure for the Arnoldi method for the calculation of a few dominant eigenvalues of a large matrix. We show that the underlying idea of implicit polynomial filtering can be utilized in much the same manner via the implicitly shifted LR iteration to create a restart procedure for the non-symmetric Lanczos algorithm for eigenvalue computations, which preserves the tri-diagonal structure of the reduced matrix. 
\end{abstract}



\section{Introduction}
The Arnoldi iteration \cite{arnoldi51} is a popular Krylov space method for calculating a few eigenvalues of a large matrix. The method relies on the generation of a sequence of Krylov vectors which determine the subspace within which approximations of the eigenvalue-eigenvector pairs are obtained. Depending on the accuracy and number of eigenpair approximations needed, the Krylov space size can become exceedingly large so that  the quality of the results may be limited by the available memory. Sorensen \cite{sorensen92} introduced an elegant procedure for restarting the Arnoldi factorization based on polynomial filters, which are applied through the implicitly shifted QR iterations on the reduced Hessenberg matrix obtained through the Arnoldi method. In particular, the use of exact shifts was shown to be successful in the convergence process of the eigenspace\cite{sorensen92} of the specified eigenvalues. The method has subsequently found widespread application through the ARPACK library \cite{arpack98}. An alternative, more modern method for eigenvalue computations is the Krylov-Schur method introduced by Stewart \cite{stewart02}, which can be found implemented in the SLEPc library \cite{slepc}. Krylov-Schur restarts are known to be less sensitive than the restarting based on the implicitly shifted QR method. However, they do not preserve the Hessenberg structure of the reduced matrix which may sometimes be required. The use of QR iterations ensures that the reduced matrix preserves its Hessenberg structure through the transforms that make up the restart process. If the underlying matrix is symmetric, the Arnoldi iteration reduces to the Lanczos algorithm, and the Hessenberg matrix reduces to a symmetric tridiagonal matrix. The QR iteration preserves the symmetric tridiagonal structure as well and, as pointed out by Sorensen \cite{sorensen92}, the implicit restart process applies equally well for the Lanczos method for symmetric matrices.  

The Lanczos algorithm introduced in \cite{lanczos50} is in fact the predecessor of the Arnoldi method and can be used for non-symmetric matrices as well. In such cases it is referred to as the non-symmetric or the bi-Lanczos. It has been used in investigationss for plasma instabilities \cite{parlett85}, linear systems solvers \cite{gutknecht97}, model reduction \cite{papakos03}, and more recently for non-linear eigenvalue problems \cite{gaaf17}. For the non-symmetric Lanczos one builds two Krylov subspaces referred to as the right and the left Krylov subspaces. The idea behind the two methods is similar, which is to obtain a projection of the original large matrix on to an appropriate reduced subspace such that, the eigenvalues may be approximated via the eigenvalues of the reduced operator. The difference being that in the Arnoldi method one obtains an orthogonal projection on to a subspace while, in the Lanczos method one obtains an oblique projection. One would therefore like to extend the idea of implicit restarts to the non-symmetric Lanczos algorithm as well. However, the reduced matrix that one obtains in such a case is a non-symmetric tridiagonal matrix, with the tridiagonal structure being the result of the recurrence relations of the Lanczos algorithm \cite{saad82}. Since the QR iterations do not preserve the banded structure of non-symmetric matrices, a straightforward application of the restart procedure put forward by Sorensen will lead to a loss of this tridiagonal structure (the Hessenberg structure will still be preserved). This loss of structure can be circumvented if one looks to the predecessor of the QR algorithm namely, the LR algorithm proposed by Rutishauser \cite{rutishauser58,rutishauser63,rutishauser91}, which has the attractive property of preserving the band structure of a matrix. This property was already pointed out by Rutishauser in \cite{rutishauser58} where the banded matrices were referred to as striped matrices. As we will show in the next section, the shifted LR transform is the appropriate analogue of the restart procedure in the case of non-symmetric Lanczos iteration. The process would necessarily require refining both the right as well as the left Krylov spaces simultaneously.

The rest of the paper is organized as follows. In section~\ref{sec:nonsymmetric_lanczos} we start with the introduction of the non-symmetric Lanczos iteration and then develop the restart procedure. In section~\ref{sec:computational_results} we apply the restart process to the Grcar matrix, and make some concluding remarks in section~\ref{sec:conclusion}.

\section{Non-symmetric Lanczos}
\label{sec:nonsymmetric_lanczos}
Lanczos first introduced his algorithm in \cite{lanczos50} as a method for tridiagonalizing a matrix, but also realized that the method could be used iteratively to find eigenvalues. For an arbitry matrix $A$, the method generates a pair of Krylov subspaces $\{v_{1},\ldots,v_{m}\}$ and $\{w_{1},\ldots,w_{m}\}$, through repeated action of $A$ and $A^{H}$ respectively. We refer to these as the right and left Krylov spaces respectively and they satisfy the biorthogonality relation $w_{i}^{H}v_{j}=\delta_{ij}$. The two subspaces are generated through a three term recurrence relation which, for a Krylov space of size $m$, can be written in matrix form as 
\begin{subequations}
	\begin{eqnarray}
		AV_{m} - V_{m}T_{m} &=& \widetilde{v}_{m+1}e_{m}^{T}, \label{eqn:lanczos_right} \\
		A^{H}W_{m} - W_{m}T^{H}_{m} &=& \widetilde{w}_{m+1}e_{m}^{T}, \label{eqn:lanczos_left} \\
		W_{m}^{H}V_{m}	&=& I_{m}, \label{eqn:biorthogonality}
	\end{eqnarray}
\end{subequations}
where, $I_{m}$ represents the Identity matrix of size $m$, $T_{m}$ is a tri-diagonal matrix of size $m$ with $T_{m}^{H}$ it's Hermitian conjugate, and $e_{m}$ is the standard unit vector. $\widetilde{v}_{m+1}$ and $\widetilde{w}_{m+1}$ represent the residual vectors at the $m^{th}$ step. If either $\widetilde{v}_{m+1}$ or $\widetilde{w}_{m+1}$ vanishes it represents the convergence of the right or the left Krylov subspaces to an invariant subspace of dimension $m$. A more serious breakdown occurs if $\widetilde{w}_{m+1}^{H}\widetilde{v}_{m+1} = 0$ with both $\widetilde{v}_{m+1}\ne0$ and $\widetilde{w}_{m+1}\ne0$, in which case a look-ahead strategy may be employed. We do not address the issues with breakdown here since it is not specifically related to the restart procedure. We refer the reader to \cite{parlett85} for the look-ahead Lanczos and to \cite{gutknecht97} for a comprehensive overview on Lanczos type solvers and the related issues of breakdown. 

As Sorensen points out for the Arnoldi method \cite{sorensen92}, if one is interested in an invariant subspace of dimension $m$, the starting vector of the Krylov subspace  must not contain components of  the generator of a cyclic subspace of dimension greater than $m$. This applies equally for the right and left Krylov subspaces generated through the Lanczos recurrence relations. Hence a non-vanishing $\widetilde{v}_{m+1}$ (respectively $\widetilde{w}_{m+1}$) implies that $v_{1}$ (respectively $w_{1}$) contains components of an invariant subspace of dimension greater than $m$. The idea behind restarts then is to discard the components of the starting vector $v_{1}$ (and $w_{1}$) along the unwanted dimensions, such that each restart process moves the Krylov space(s) closer to being invariant. For the Arnoldi method Sorensen \cite{sorensen92} proposed to achieve this via polynomial filtering, \textit{i.e.} replacing
\begin{subequations}
	\begin{eqnarray}
		v_{1} \leftarrow \psi(A)v_{1}, \\	
		\psi(\lambda) = (1/\tau)\Pi_{j=1}^{p}(\lambda - \mu_{j}).
	\end{eqnarray}
\end{subequations}
Obviously $\psi(\lambda)$ is the filtering polynomial, $\tau$ is a normalization constant and each $\mu_{j}$ specifies a node of the polynomial. The polynomial acts on $v_{1}$ to filter out the part of the spectrum of $A$ that is close to each $\mu_{j}$. If a particular $\mu_{j}$ corresponds to an exact eigenvalue of $A$, then components of the corresponding eigenvector are completely filtered out from $v_{1}$ (at least in exact arithmetic). 
The node $\mu_{j}$ is referred to as a shift since the application of the polynomial filtering relies on the shifted QR algorithm, where $\mu_{j}$ is used as the shift. As shown below for the case of a single shift, an analogous procedure can be followed using a shifted LR algorithm which achieves the same effect of applying a polynomial filter to the starting vector $v_{1}$. Starting with the Lanczos relation for the right subspace \eqref{eqn:lanczos_right}, and adding and subtracting $\mu V_{m}$ we obtain
\begin{subequations}
	\begin{eqnarray}
		(A - \mu I) V_{m} - V_{m}(T_{m} - \mu I) &=& \widetilde{v}_{m+1}e_{m}^{T} \label{alg:shifted_lr_right_1}\\
		(A - \mu I) V_{m} - V_{m}(L_{1}R_{1}) &=& \widetilde{v}_{m+1}e_{m}^{T} \label{alg:shifted_lr_right_2}\\
		(A - \mu I) V_{m}L_{1} - V_{m}(L_{1}R_{1})L_{1} &=& \widetilde{v}_{m+1}e_{m}^{T}L_{1} \label{alg:shifted_lr_right_3}\\
		A(V_{m}L_{1}) - (V_{m}L_{1})(R_{1}L_{1} + \mu I) &=& \widetilde{v}_{m+1}e_{m}^{T}L_{1}	\label{alg:shifted_lr_right_4} \\
		AV'_{m} - V'_{m}T'_{m} &=& \widetilde{v}_{m+1}e_{m}^{T}L_{1}	\label{alg:shifted_lr_right_5} .
	\end{eqnarray}	
\end{subequations}
Here we have set $V'_{m} = V_{m}L_{1}$ and $T'_{m} = (R_{1}L_{1} + \mu I)$. The matrices $L_{1}, R_{1}$ are the lower and upper triangular matrices obtained from the LU decomposition of $(T_{m} - \mu I)$. The matrix $L_{1}$ can be required to be unit triangular (all entries on the main diagonal are ones), in which case the LU decomposition is unique. Furthermore, $L_{1}$ for a tridiagonal matrix only consists of one sub-diagonal (in addition to the main diagonal). One can easily recognize that the new reduced matrix $T'_{m}$ is a result of one step of the shifted LR iteration. Hence $T'_{m}$ retains the tridiagonal structure of the $T_{m}$ \cite{rutishauser58,rutishauser91}. The relationship between starting vectors of the two spaces $V_{m}$ and $V'_{m}$ can be obtained by multiplying \eqref{alg:shifted_lr_right_2} by $e_{1}$, \textit{i.e.}
\begin{eqnarray}
		(A - \mu I) V_{m}e_{1} - (V'_{m})R_{1}e_{1} &=& \widetilde{v}_{m+1}e_{m}^{T}e_{1} \nonumber \\
		\implies (A - \mu I) v_{1}  &=& v'_{1}\rho_{11} \nonumber,
\end{eqnarray}
where $\rho_{11}=e^{T}_{1}R_{1}e_{1}$. This clearly shows the filtering operation done on the original vector $v_{1}$ to generate the new vector $v'_{1}$. 

Since the Lanczos method creates a biorthogonal basis, one must simultaneously transform the left basis $W_{m}$ to maintain the biorthogonality property. It is easy to see that the necessary transform to maintain biorthogonality is $W'_{m} = W_{m}L_{1}^{-H}$, since,
\begin{eqnarray}
	W'^{H}V'_{m} = (W_{m}L_{1}^{-H})^{H}(V_{m}L_{1}) = L_{1}^{-1}(W^{H}_{m}V_{m})L_{1} = I \nonumber
\end{eqnarray}
Substituting the relation obtained from the LU decomposition $T_{m}^{H} = (R_{1}^{H}L_{1}^{H} + \bar{\mu}I)$ in to \refeqn{eqn:lanczos_left} we can obtain the modified Lanczos relation for the left Krylov space as
\begin{subequations}
	\begin{eqnarray}
		A^{H}W_{m}	- W_{m}(R^{H}_{1}L^{H}_{1} + \bar{\mu}I) &=& \widetilde{w}_{m+1}e_{m}^{T} \label{alg:transform_left_1} \\
		A^{H}(W_{m}L_{1}^{-H})	- W_{m}(L_{1}^{-H}L_{1}^{H})(R^{H}_{1}L^{H}_{1} + \bar{\mu}I)L_{1}^{-H} &=& \widetilde{w}_{m+1}e_{m}^{T}L_{1}^{-H} \label{alg:transform_left_2} \\
		A^{H}(W_{m}L_{1}^{-H})	- (W_{m}L_{1}^{-H})(L^{H}_{1}R^{H}_{1} + \bar{\mu}I) &=& \widetilde{w}_{m+1}e_{m}^{T}L_{1}^{-H} \label{alg:transform_left_3} \\
		A^{H}W'_{m}	- W'_{m}(T'_{m})^{H} &=& \widetilde{w}_{m+1}e_{m}^{T}L_{1}^{-H} \label{alg:transform_left_4}.
	\end{eqnarray}
\end{subequations}
Conveniently the structure of the Lanczos iteration for the left Krylov space is also preserved. Noting that $L_{1}^{-H}$ is upper triangular, one can again expose the relationship between the generating vectors of the two left Krylov spaces as
\begin{eqnarray}
	W'e_{1} &=& (W_{m}L_{1}^{-H})e_{1} \nonumber \\
	\implies w'_{1} &=& w_{1}(e_{1}^{T}L_{1}^{-H}e_{1}) \nonumber.
\end{eqnarray}
Clearly $w'_{1}$ is simply a scalar multiple of the old vector $w_{1}$ and no filtering of the generating vector has occurred. In order to ensure that we filter the left Krylov space as well, we perform one step of the shifted LR iteration with the conjugated shift $\bar{\mu}$ on the reduced matrix $(T'_{m})^{H}$ obtained in \refeqn{alg:transform_left_4}. 
\begin{subequations}
	\begin{eqnarray}
		(A - \mu I)^{H} W'_{m} - W'_{m}(T'_{m} - \mu I)^{H} &=& \widetilde{w}_{m+1}e_{m}^{T}L_{1}^{-H}, \label{alg:shifted_lr_left_1}\\
		(A - \mu I)^{H} W'_{m} - W'_{m}(L_{2}R_{2}) &=& \widetilde{w}_{m+1}e_{m}^{T}L_{1}^{-H}, \label{alg:shifted_lr_left_2}\\
		(A - \mu I)^{H} W'_{m}L_{2} - W'_{m}(L_{2}R_{2})L_{2} &=& \widetilde{w}_{m+1}e_{m}^{T}L_{1}^{-H}L_{2}, \label{alg:shifted_lr_left_3} \\
		A(W'_{m}L_{2}) - (W'_{m}L_{2})(R_{2}L_{2} + \bar{\mu}I) &=& \widetilde{w}_{m+1}e_{m}^{T}L_{1}^{-H}L_{2},	\label{alg:shifted_lr_left_4} \\
		AW''_{m} - W''_{m}(T''_{m})^{H} &=& \widetilde{w}_{m+1}e_{m}^{T}L_{1}^{-H}L_{2}.	\label{alg:shifted_lr_left_5}
	\end{eqnarray}	
\end{subequations}
One may again obtain the relation between the starting vectors of the two spaces as 
\begin{eqnarray}
	(A^{H} - \bar{\mu}I)w'_{1} = w''_{1}(e_{1}^{T}R_{2}e_{1}),
\end{eqnarray}
which clearly shows the filtering operation performed on the starting vector of the left Krylov space. Obviously the appropriate transform for the right subspace to maintain orthogonality is $V''_{m} = V'_{m}L_{2}^{-H} = V_{m}L_{1}L_{2}^{-H}$. Again, note that the upper triangular $L_{2}^{-H}$ implies that the new $v''_{1}$ is simply the scalar multiple of $v'_{1}$ and no additional filtering occurs for $v_{1}$ in this step. One can write the corresponding modified Lanczos relation as
\begin{eqnarray}
	AV''_{m} - V''_{m}T''_{m} &=& \widetilde{v}_{m+1}e_{m}^{T}L_{1}L_{2}^{-H}. \label{alg:transform_right_1}
\end{eqnarray}

The above process can be repeated for $p$ unwanted shifts. We denote by $L_{1}^{p} = L_{11}L_{12}\ldots L_{1p} $ as the product of the lower triangular matrices generated due to $p$ shifted-LR steps for the right Krylov space $V_{m}$, and by $L^{p}_{2}=L_{21}L_{22}\ldots L_{2p}$ as the product of the lower triangular matrices due to the $p$ shifted-LR iterations for the left Krylov space. Then for a Krylov space size of $m=k+p$ we have two modified Lanczos relations
\begin{subequations}
\begin{eqnarray}
		AV''_{k+p} - V''_{k+p}T''_{k+p} &=& \widetilde{v}_{k+p+1}e_{k+p}^{T}L_{1}^{p}(L_{2}^{p})^{-H}, \label{eqn:modified_lanczos_right_1} \\
		A^{H}W''_{k+p} - W''_{k+p}(T''_{k+p})^{H} &=& \widetilde{w}_{k+p+1}e_{k+p}^{T}(L_{1}^{p})^{-H}L_{2}^{p}	\label{eqn:modified_lanczos_left_1}.		
\end{eqnarray}
\end{subequations}

We may take a closer look at the structure of the residual matrices on the right hand side of \refeqn{eqn:modified_lanczos_right_1}. $L_{1}^{p}$ is a product of $p$ matrices that are lower triangular with just one subdiagonal. $L_{1}^{p}$ then is lower triangular with $p$ non-zero subdiagonals. $(L_{2}^{p})^{-H}$ is upper triangular and the product $L_{1}^{p}(L_{2}^{p})^{-H}$ therefore has $p$ non-zero subdiagonals. Left multiplication by $e_{k+p}^{T}$ therefore has the form
\begin{eqnarray}
	e_{k+p}^{T}L_{1}^{p}(L_{2}^{p})^{-H} = (\underbrace{0,0\ldots,\theta_{k+p}}_{k},\underbrace{b_{1}^{T}}_{p} ) \nonumber
\end{eqnarray}
where, $\theta_{k+p} = e_{k+p}^{T}(L_{1}^{p}(L_{2}^{p})^{-H})e_{k}$. Therefore the matrix on the right hand side of \eqref{eqn:modified_lanczos_right_1} has zeros in the first $k-1$ columns and the $k^{th}$ column is simply $\theta_{k+p}\widetilde{v}_{k+p+1}$. The remaining columns are non-zero in general. A very similar structure is obtained for the residual matrix in the right hand side of \refeqn{eqn:modified_lanczos_left_1} with zeros in the first $k-1$ columns and the $k^{th}$ column being equal to $\phi_{k+p}\widetilde{w}_{k+p+1}$, with $\phi_{k+p} = e_{k+p}^{T}((L_{1}^{p})^{-H}L_{2}^{p})e_{k}$ 

Partitioning the matrices such that
\begin{subequations}
	\begin{eqnarray}
			V''_{k+p} = (V''_{k},V''_{p}), & \hspace{10pt} T''_{k+p} = \begin{pmatrix}
				T''_{k} 	& \delta_{k+1}e_{k}e_{1}^{T} \\
				\beta_{k+1}e_{1}e_{k}^T	&	T''_{p}
			\end{pmatrix}, \nonumber\\
			W''_{k+p} = (W''_{k},W''_{p}), & \hspace{10pt} (T''_{k+p})^{H} = \begin{pmatrix}
			(T''_{k})^{H} 	& \bar{\beta}_{k+1}e_{k}e_{1}^{T} \\
			\bar{\delta}_{k+1}e_{1}e_{k}^T	&	(T''_{p})^{H}
		\end{pmatrix}, \nonumber
	\end{eqnarray}
\end{subequations}
with the length of the $e_{i}$ vectors understood to be such that the resulting matrices are consistent. We can write the modified Lanczos relations of \eqref{eqn:modified_lanczos_right_1} and \eqref{eqn:modified_lanczos_left_1} as
\begin{eqnarray}
	A(V''_{k},V''_{p}) = (V''_{k},V''_{p}) \begin{pmatrix}
		T''_{k} 	& \delta_{k+1}e_{k}e_{1}^{T} \\
		\beta_{k+1}e_{1}e_{k}^T	&	T''_{p}
	\end{pmatrix} + \begin{pmatrix}
	\theta_{k+p}\widetilde{v}_{k+p+1}e_{k}^{T}, M_{v}
\end{pmatrix} \label{eqn:modified_lanczos_left_2}, \nonumber \\
	A^{H}(W''_{k},W''_{p}) = (W''_{k},W''_{p}) \begin{pmatrix}
		(T''_{k})^{H} 	& \bar{\beta}_{k+1}e_{k}e_{1}^{T} \\
		\bar{\delta}_{k+1}e_{1}e_{k}^T	&	(T''_{p})^{H}
	\end{pmatrix} + \begin{pmatrix}
	\phi_{k+p}\widetilde{w}_{k+p+1}e_{k}^{T}, M_{w}
\end{pmatrix} \label{eqn:modified_lanczos_right_2}, \nonumber
\end{eqnarray}
Finally, equating the individual columns on both sides and discarding columns $k+1,\ldots,k+p$ we are left with the new Krylov spaces of order $k$ and the Lanczos relations
\begin{subequations}
	\begin{eqnarray}
			AV''_{k} - V''_{k}T''_{k} &=& \widetilde{v}''_{k+1}e_{k}^{T} \label{eqn:restarted_lanczos_right_1}, \\
			A^{H}W''_{k} - W''_{k}(T''_{k})^{H} &=& \widetilde{w}''_{k+1}e_{k}^{T} \label{eqn:restarted_lanczos_left_1}, \\
			(W''_{k})^{H}V_{k} &=& I \label{eqn:orthogonality_new}.
	\end{eqnarray}
\end{subequations}
The new residual vectors are defined as
\begin{subequations}
	\begin{eqnarray}
		\widetilde{v}''_{k+1}	&=&	\beta_{k+1}V''_{p}e_{1} + \theta_{k+p}\widetilde{v}_{k+p+1}, \label{eqn:residual_new_right} \\
		\widetilde{w}''_{k+1}	&=&	\bar{\delta}_{k+1}W''_{p}e_{1} + \phi_{k+p}\widetilde{w}_{k+p+1}, \label{eqn:residual_new_left} 		
	\end{eqnarray}
\end{subequations}
which may be normalized appropriately such that the inner product of the new Krylov vectors is unity. The Lanczos process may now be carried out again to generate the next $p$ vectors of the right and left Krylov spaces and the cycle may be repeated till an adequately converged subspace has been obtained. In the case that exact shifts (eigenvalues of $T_{m}$) are used for the restart procedure $\beta_{k+1}$ and $\delta_{k+1}$ are both zero in exact arithmetic.  

As a final note, we mention that Della-Dora \cite{delladora75} introduced a class of algorithms of the GR type of which, LR and QR are special cases. Watkins then introduced generic bulge-chasing algorithms for the entire GR class of methods\cite{watkins91}. The shifted LR algorithm used in the restart procedure outlined above can therefore be carried out in an implicit manner through the bulge-chase sequence of Watkins \cite{watkins91}. For real matrices, the operations can be confined in the real space by using the double-shift strategy introduced by Francis \cite{francis62}. The entire restart process is put together in algorithm~\ref{alg:restarted_lanczos}.

\begin{algorithm}[h]
\caption{Restarted nonsymmetric Lanczos}
\label{alg:restarted_lanczos}
\textbf{Input: } $V_{k+p}, \widetilde{v}_{k+p+1}, W_{k+p}, \widetilde{w}_{k+p+1}, T_{k+p}$ \\
\textbf{Input:} $\mu_{1},\mu_{2},\ldots\mu_{p}$	 \Comment{Unwanted shifts} \\
\textbf{Output: } $V_{k}, \widetilde{v}_{k+1}, W_{k}, \widetilde{w}_{k+1}, T_{k}$
\begin{algorithmic}[1]
\Require{$AV_{k+p} - V_{k+p}T_{k+p} = \widetilde{v}_{k+p+1}e_{k+p}^T$} \Comment{Right Lanczos relation}
\Require{$A^{H}W_{k+p} - W_{k+p}T^{H}_{k+p} = \widetilde{w}_{k+p+1}e_{k+p}^T$} \Comment{Left Lanczos relation}
\Require{$W_{k+p}^{H}V_{k+p} = I$; $W_{k+p}^{H}\widetilde{v}_{k+p+1} = 0$; $V_{k+p}^{H}\widetilde{w}_{k+p+1} = 0$}; \Comment{Bi-orthogonality}
\Procedure{Refine right subspace}{}
	 \State $L_{r} \gets I_{k+p}$
      \For{$j \gets 1$ to $p$}
			\State $T_{k+p} \gets L^{-1}T_{k+p}L$	\Comment{Implicitly shifted LR with shift $\mu_{j}$}
			\State $L_{r} \gets LL_{r}$
\EndFor
\EndProcedure
\Procedure{Refine left subspace}{}
\State $L_{l} \gets I_{k+p}$
\For{$j \gets 1$ to $p$}
\State $T^{H}_{k+p} \gets L^{-1}T^{H}_{k+p}L$	\Comment{Implicitly shifted LR with shift $\bar{\mu}_{j}$}
\State $L_{l} \gets LL_{l}$
\EndFor
\EndProcedure
\Procedure{Update}{}
\State  $\beta \gets e^{T}_{k+1}T_{k+p}e_{k}$;\hspace{20pt} $\delta \gets e^{T}_{k}T_{k+p}e_{k+1}$
\State  $\theta = e_{k+p}^{T}(L_{r}(L_{l})^{-H})e_{k}$; \hspace{20pt} $\phi = e_{k+p}^{T}(L_{r}^{-H}L_{l})e_{k}$
\State $\widetilde{v}_{k+1}	\gets	\beta V_{k+p}e_{k+1} + \theta \widetilde{v}_{k+p+1}$ \Comment{New right residual}
\State $\widetilde{w}_{k+1}	\gets	\bar{\delta} W_{k+p}e_{k+1} + \phi \widetilde{w}_{k+p+1}$ \Comment{New left residual}
\State $V_{k} \gets V_{k+p}(L_{r}L_{l}^{-H})\begin{pmatrix}
	I_{k}\\0_{p}
\end{pmatrix}$ \Comment{New right Krylov space}
\State $W_{k} \gets W_{k+p}(L_{r}^{-H}L_{l})\begin{pmatrix}
	I_{k}\\0_{p}
\end{pmatrix}$ \Comment{New left Krylov space}
\State $T_{k} \gets \begin{pmatrix}
	I_{k} & 0_{p}
\end{pmatrix} T_{k+p}\begin{pmatrix}
	I_{k}\\0_{p}
\end{pmatrix}$ 	\Comment{New tridiagonal matrix}
\EndProcedure \\
\Return $V_{k}, \widetilde{v}_{k+1}, W_{k}, \widetilde{w}_{k+1}, T_{k}$
\end{algorithmic}
\end{algorithm}
\FloatBarrier

\section{Computational Results}
\label{sec:computational_results}
We present the results of computational tests performed in the Grcar matrix which is highly non-normal and has presented problems with convergence in previous studies. In his implicit restart work Sorensen \cite{sorensen92} indeed points out that the restarted Arnoldi has trouble converging to the left-most part of the spectrum and even though the algorithm claimed convergence, what was obtained was in fact part of the pseudospectrum. 

\begin{figure}[]
	\centering
	\includegraphics[width=0.75\textwidth]{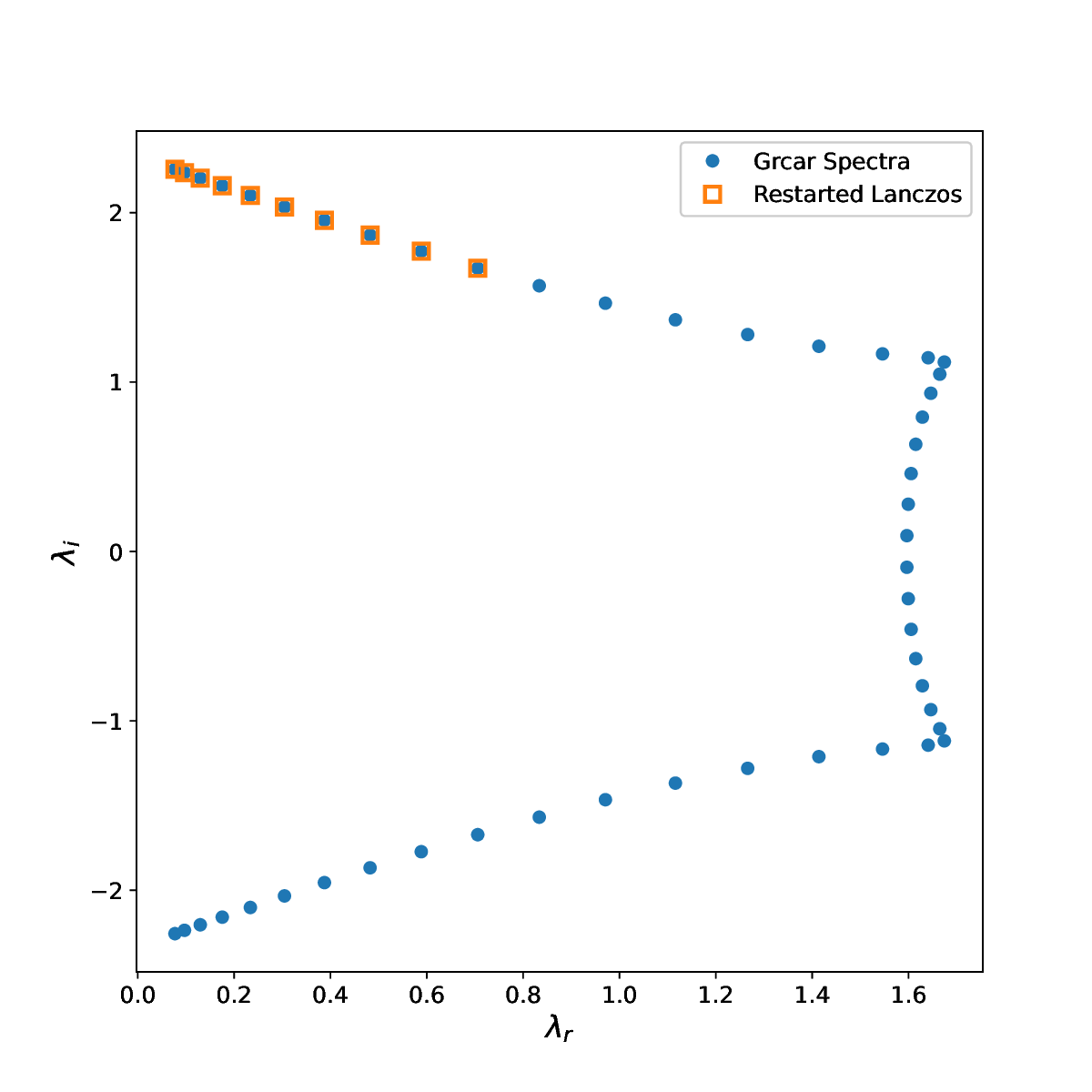}
	\caption{Spectrum obtained from the restarted Lanczos after $11$ restart steps.}
	\label{fig:grcar_eigs}
	\includegraphics[width=0.75\textwidth]{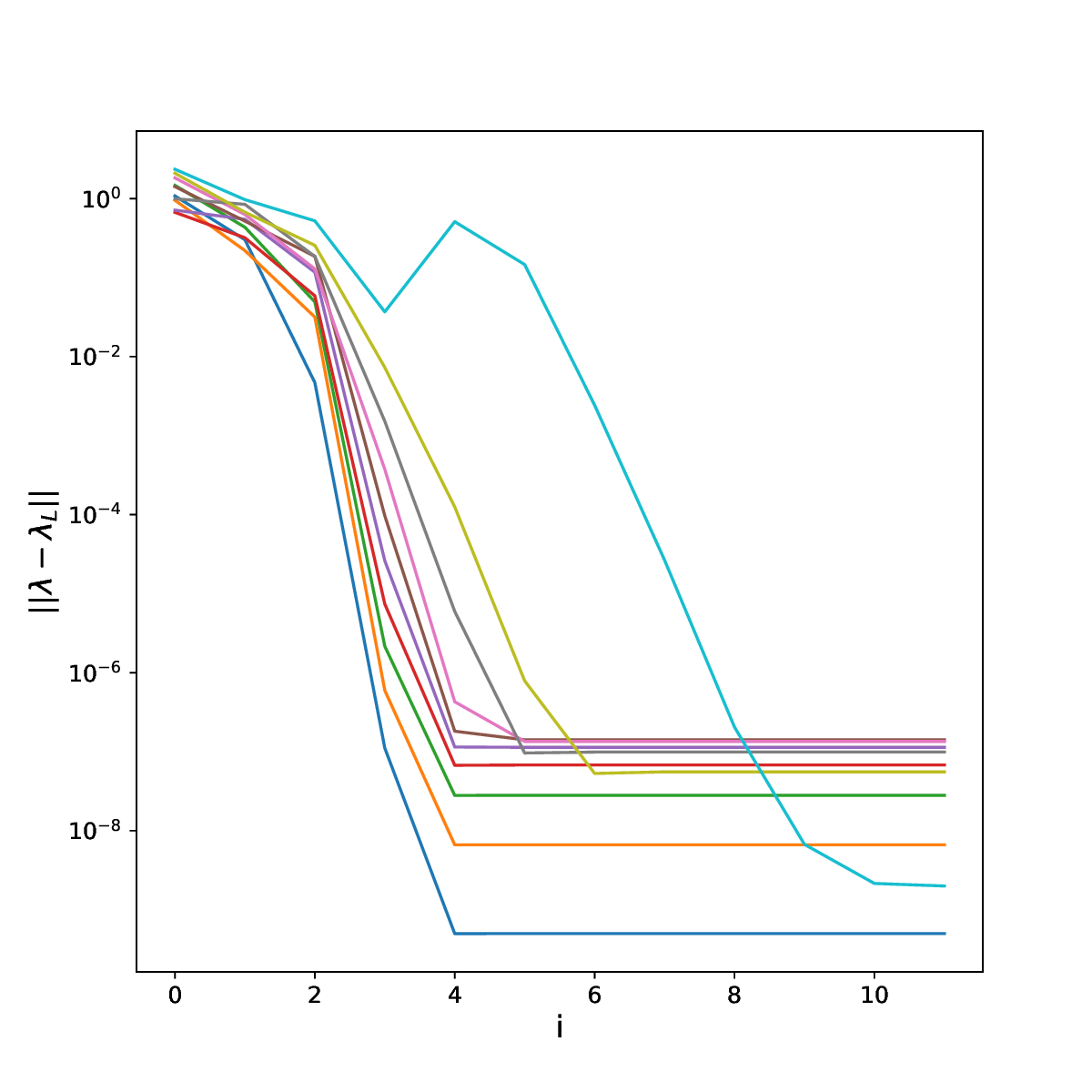}
	\caption{Error in the eigenvalues after each restart $i$.}
	\label{fig:eig_error}
\end{figure}

We present the results of the restarted Lanczos method applied to the $50\times50$ Grcar matrix for $k=10$ and $p=10$.  Figure~\ref{fig:grcar_eigs} shows the spectrum obtained after performing eleven restarts of the Lanczos algorithm. The restarts were performed such that the $k$ eigenvalues with the largest imaginary part were retained for the reduced operator and the remaining were used as shifts to be discarded from the subspace. The same part of the spectrum was sought by Sorensen in \cite{sorensen92}. In this particular case we obtained $\widetilde{w}_{m+1}^{T}\widetilde{v}_{m+1} \sim O(10^{-11})$ even though the individual residuals were both of order $10^{-3}-10^{-4}$. At this particular moment, one would need to employ a look-ahead step of the Lanczos however, we have shown the spectrum of the reduced matrix at this point. One may think of the spectrum to have converged to $O(10^{-4})$. As seen from the figure, the spectrum of the reduced operator matches quite well with the original spectrum of the Grcar matrix. In figure~\ref{fig:eig_error}, we show the convergence history of the individual eigenvalues with each restart step as the Lanczos algorithm progresses. Clearly the restart process is working well to shift the Krylov spaces towards the wanted region of the spectrum. The error in the eigenvalues is of $O(10^{-7})$ even though the perturbation to the reduced matrix is of order $10^{-4}$. This is in contrast to the results obtained in \cite{sorensen92} which reported large perturbations to the eigenvalues even when the Arnoldi method reported convergence. We suspect this is due to the fact that the non-symmetric Lanczos approximates both the right and left eigenspaces simultaneously and leads to a lower error in the truncated matrix. We expect this to be particularly useful in hydrodynamic problems where highly non-normal matrices are a routine occurrence and where problems with convergence of the spectrum have often been reported (see \cite{rahkola17} for example). 

At this point we make a note of caution that for large Grcar matrices our attempts have been somewhat less successful and the convergence sometimes fails. We expect this is due to the naive implementations of the implicit LR method that we have done in Julia where, the standard checks for small sub-diagonal elements have not been performed. In our investigations of such cases we indeed do find small sub diagonal elements which lead to loss of precision. Very small sub diagonal elements also lead to rapid loss of biorthogonality of the two subspaces. Similarly we have not paid attention to the issue arising out of the Lanczos algorithm itself, except for employing a double (two-sided) Gram-Schmidt to ensure biorthogonality. We also note that for large number of restarts, the biorthogonality property of the two subspaces is progressively lost, probably due to accruing floating point errors. Hence we expect some method of reorthogonalization would be required to ensure stability for very long calculations, which has not been done in the current work. The (implicit) LR decomposition is not unique. Uniqueness is ensured for unit main diagonal of the lower triangular matrix however, this does not pay any heed to conditioning of the transforming matrices. A better strategy of building the matrices could be pursued which has better conditioning while at the same time preserves the tridiagonal structure. These issues would require careful implementation of all the individual components and we do not address those in the current work. 

Finally we note that very similar work has been reported in \cite{grimme96} where hyperbolic transforms (HR) are used for restarts of the bi-Lanczos method, and in \cite{grimme94} where the LR transformations are used, albeit in the context of model reduction. 

\section{Conclusion}
\label{sec:conclusion}
We present an algorithm to restart the non-symmetric Lanczos method which is based on the idea of polynomial filtering via the implicit QR method proposed by Sorensen \cite{sorensen92} for restarting the Arnoldi iteration. The (implicitly) shifted LR method is shown to be the appropriate analogue for restarting the non-symmetric Lanczos algorithm for structure preserving restarts. It is shown that the polynomial filtering process needs to be carried out for both the right and the left Krylov spaces and the appropriate transforms for maintaining biorthogonality of the two spaces are highlighted. Computational results are shown for a Grcar matrix which is known to be highly non-normal and the spectrum is found to converge adequately even when the residual error is relatively large. 

\section*{Acknowledgments}
The authors would like to thank Professor Elias Jarlebring for his helpful comments on the manuscript. The authors acknowledge support of Nordita and the Swedish Research Council Grant No. 2018-04290. Nordita is partially supported by Nordforsk.


\FloatBarrier

\bibliographystyle{siamplain}
\bibliography{references}
\end{document}